\documentclass[runningheads]{llncs}
\usepackage[utf8]{inputenc}

\usepackage[utf8]{inputenc}
\usepackage{graphicx}
\usepackage{amsmath}
\usepackage{hyperref}
\usepackage{subfig}
\usepackage{caption}
\usepackage[table]{xcolor}
\usepackage{float}
\usepackage{paralist}
\usepackage{url}
\usepackage{tabto}
\usepackage{latexsym}
\usepackage{textcomp}
\usepackage{amsfonts}

\begin{document}
\title{MapSDI: A Scaled-up Semantic Data Integration Framework for Knowledge Graph Creation}
\titlerunning{MapSDI}
%
%
\author{Samaneh Jozashoori\inst{1,2}, 
        Maria-Esther Vidal\inst{2,1}}
\authorrunning{Jozashoori and Vidal}
%
\institute{L3S Research Center, Leibniz University of Hannover, Germany \\ 
\email{jozashoori@l3s.de}\\
\and
TIB Leibniz Information Center for Science and Technology, Germany\\
\email{maria.vidal@tib.eu}
}
\maketitle         
\begin{abstract}
Semantic web technologies have significantly contributed\linebreak with effective solutions for the problems of data integration and knowledge graph creation. However, with the rapid growth of big data in diverse domains, different interoperability issues still demand to be addressed, being scalability one of the main challenges.
In this paper, we address the problem of knowledge graph creation at scale and provide MapSDI, a mapping rule-based framework for optimizing semantic data integration into knowledge graphs. MapSDI allows for the semantic enrichment of large-sized, heterogeneous, and potentially low-quality data efficiently. The input of MapSDI is a set of data sources and mapping rules being generated by a mapping language such as RML. First, MapSDI pre-processes the sources based on semantic information extracted from mapping rules, by performing basic database operators; it projects out required attributes, eliminates duplicates, and selects relevant entries. All these operators are defined based on the knowledge encoded by the mapping rules which will be then used by the semantification engine (or RDFizer) to produce a knowledge graph.
We have empirically studied the impact of MapSDI on existing RDFizers, and observed that knowledge graph creation time can be reduced on average in one order of magnitude. It is also shown, theoretically, that the sources and rules transformations provided by MapSDI are data-lossless.

\end{abstract}
\section{Introduction}
\label{sec:introduction}
Knowledge graph creation as a method for knowledge representation has been through a significant progress with the development of semantic web technologies in recent years. The semantic web perspective of making the data and information more accessible to machines \cite{antoniou2004semantic} by providing a unified view of data residing in different sources with heterogeneous structures, had made semantic web technologies desirable candidates to be used in semantic data integration systems and knowledge graph creation. Coordinately, with the rapid growth of available big data in different domains, semantic data integration systems are required to be scaled up in order to transfer big data into an actionable knowledge represented in knowledge graphs.
RDF\footnote{\url{https://www.w3.org/RDF/}} or Resource Description Framework, as a standard model on the web for describing the metadata of resources, is a common data model to create linked data and knowledge graphs. Nevertheless, in many domains such as biomedicine and biology, a massive amount of generated big data is not available in this format. To create a knowledge graph from non-RDF big data sources, it is required to define mapping rules for data model transformation along with semantic data integration. However, to scale up to big data, RDFizers need to be empowered with efficient processes for removing duplicates, and projecting and selecting only relevant attributes and data.

\noindent \textbf{Problem and Objective:} 
We tackle the problem of semantic big data integration into a knowledge graph and focus on scalability issues present in existing mapping rule-based RDFizers. As proof of concept, we concentrate on RML \cite{dimou2013extending}, a mapping language that expresses mappings from hierarchical sources into a RDF graph, and the RMLmapper and SDM-RDFizer as engines for RML triple maps. We show how dominant dimensions of big data, e.g., volume, variety, and veracity, negatively impact on the performance of these two engines and prevent them from scaling up to large datasets composed of duplicated data. 

\noindent \textbf{Our Proposed Approach:} The main idea of this article is to present MapSDI, a framework for transforming big data into a knowledge graph. As traditional frameworks for knowledge graph creation, MapSDI resorts to semantification engines for creating RDF triples; however, to minimize the impact of big data dimensions, MapSDI performs transformations in the input datasets to eliminate irrelevant attributes and duplicates. MapSDI is able to exploit knowledge encoded in the triple maps to determine which attributes and data are required. It also falls back on well-known properties of the relational algebra operators, e.g., pushing down of the projections and selections, in order to pre-process the input datasets before the mappings are executed. First, by projecting out the attributes that are mentioned in a mapping rule, duplicates are eliminated and the size of the input data is reduced. Similarly, the projection of attributes positively impacts on the performance of joins between triple maps. We have empirically studied the performance of MapSDI framework on a testbed of real-world datasets. Observed results suggest that MapSDI framework is able to empower the performance of the studied RDFizers, reducing the semantification time by up to one order of magnitude (on average). While, we show theoretically that mentioned pre-processing of input datasets does not lead to any data lossness in the output i.e., generated knowledge graph remains the same.
\noindent \textbf{Contributions:}
The main contribution of this work is MapSDI, a framework able to pre-process big datasets with the aim of empowering scalability of existing RDFizers. Another important contribution represents both theoretical and empirical evaluation of the effect of the MapSDI framework on the tasks of knowledge graph creation; the testbeds are defined over real-world datasets of genomic data and show the benefits of the pre-processing step in the MapSDI framework.

This article is structured as follows:
\autoref{sec:motivating} motivates the problem of semantic data integration over a set of biomedical datasets, \autoref{sec:framework} describes the MapSDI framework, the main transformation rules and their correctness, and \autoref{sec:experiments} reports on the results of the empirical study. Related work is presented in \autoref{sec:relatedWork}, and finally, \autoref{sec:conclusions} concludes and give insights for future work.
\begin{figure}[h!]
\includegraphics[width=\textwidth]{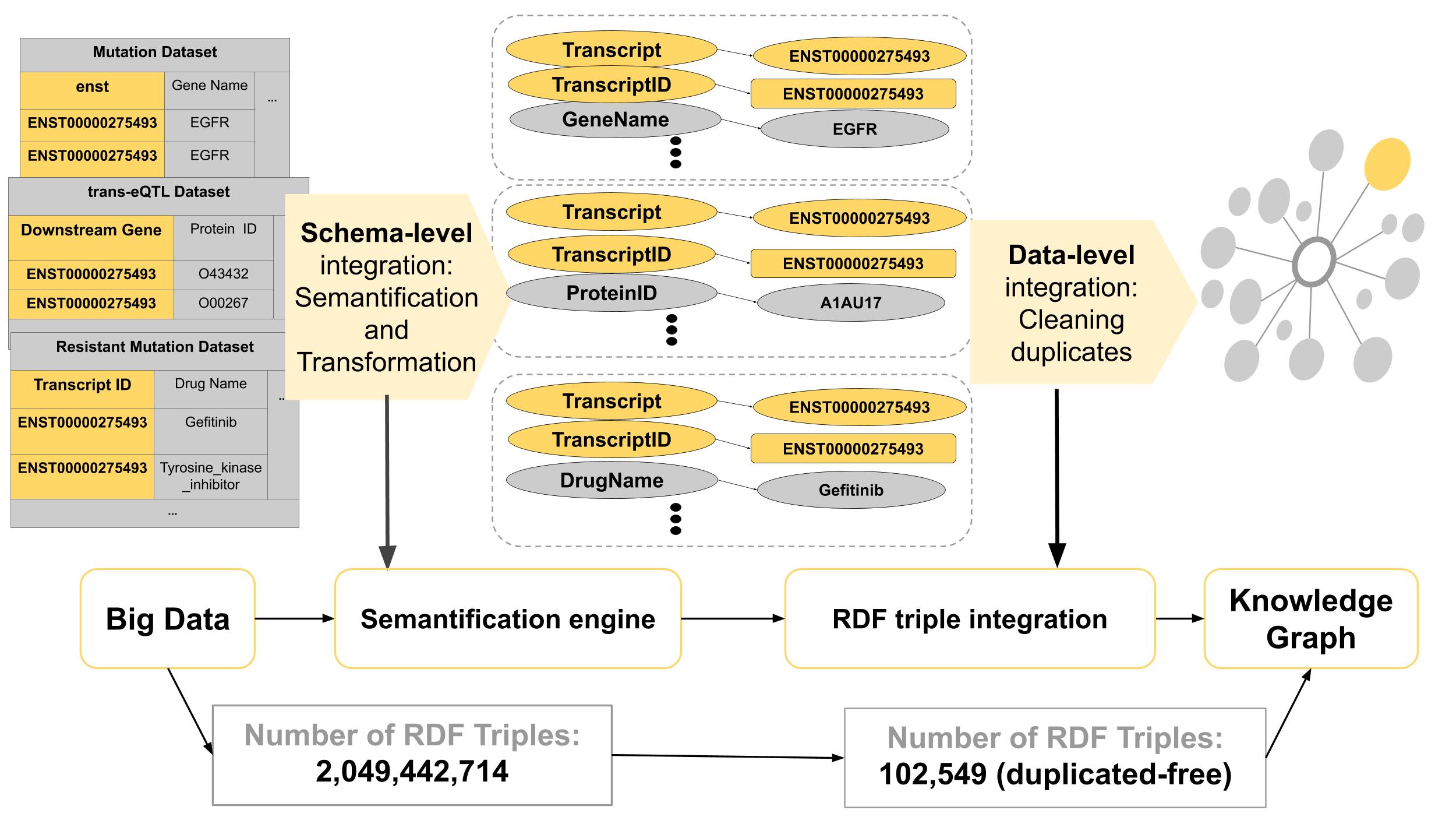}
\caption{\textbf{Motivating example}. A traditional framework where datasets characterized by big data dominant dimensions, i.e., volume, variety, and veracity, are semantically enriched and integrated into a knowledge graph. A semantification engine performs the schema-level integration by executing mapping rules, e.g., RML triple maps. Because data can be duplicated across the input datasets, a large number of RDF triples can be generated, e.g., 2,049,442,714 RDF triples. However, when duplicates are removed and cleaning techniques are performed, only 102,549 RDF triples (duplicated-free) are included in the knowledge graph.} \label{fig1}
\end{figure}
\section{Motivating Example}
\label{sec:motivating}
We motivate our work with a traditional pipeline for transforming three\linebreak datasets into instances of a knowledge graph. The datasets contain information about mutations of genes, downstream genes, and drug resistances caused by mutations. 
These files are composed of up to 39 attributes (the mutation dataset), and their sizes are 186.4 MB, 71.9 GB, and 559 KB, respectively. The semantification of these datasets just for the concept transcript is performed using three RML triple maps. These triple maps consider only the attribute that represents \emph{transcript} using a different name in each dataset (enst, downstream gene, transcript id).
This process ends up producing 2,049,442,714 RDF triples. However, because of overlaps across the three files, a large number of duplicates are generated, being reduced the output to only 102,549 duplicate-free RDF triples when cleaning and duplicate elimination are performed. \autoref{fig1} illustrates this pipeline; it receives the three datasets and outputs the RDF triples to be included in the knowledge graph. As observed, in this real-world example, the pipeline for this semantic integration task is performed via two separated steps including:
\begin{inparaenum}[(I\upshape)]
\item
\textbf{Schema-level integration}: Ontology based data semantification and mapping rule-based data transformations.
\item \textbf{Data-level integration}: Redundancy elimination and cleaning. 
\end{inparaenum}
To explain the situation reported in this example, let us consider the meaning of these three datasets. A \textit{transcript} refers to a ribonucleic acid via which a gene is expressed; it is used to synthesize a protein \cite{principles}. As it can be seen in \autoref{fig1}, \textit{transcript} as a concept, can be represented with different labels in various databases which means that it cannot be distinguished and treated as the same concept unless being semantified according to the unified schema. Therefore, the first step of integration in the framework is to unify all the concept representations residing in different datasets by defining RML triple maps while transforming the data into RDF. 
The data semantification allows for also detecting duplicated data that were not recognizable before. Consequently, in the second step, the redundant data that are now represented as RDF triples are eliminated. It should be noted that the overall number of generated triples from different sources are 16,445 times the number of non-redundant triples which means that there is a considerable amount of duplicated data that could not be detected in the raw files. Considering the fact that similarity-based comparisons between RDF triples are more expensive than between the relational data model, specifically in case of having huge amount of data, leaves room to think about providing a more efficient and low-cost approach to create knowledge graphs. In this paper, we address the problem of semantic data integration motivated in this example, and present MapSDI, a framework able to pre-process input datasets and avoid the generation of duplicated RDF triples. 
MapSDI is able to extract from the RML triple maps the knowledge required to pre-process the input datasets by means of the execution of basis relational algebra operations like the projection of attributes. Albeit simple, the transformations executed by MapSDI enable to project out only attributes that are utilized in the three triple maps, allowing the RDFizer to produce 102,549 duplicate-free RDF triples.  

\section{The MapSDI Framework}
\label{sec:framework}
The MapSDI framework relies on a data integration system $DIS_G$ which enables the transformation and integration of heterogeneous data in a knowledge graph $G$. The data integration system $DIS_G = \langle O,S,M \rangle$ is defined in terms of three components i.e., $O$ a unified schema or ontology, $S$ a set of data sources, and $M$ a set of mapping rules \cite{lenzerini2002data}.
\begin{itemize}
    \item [$\bullet$] The unified schema $O$ is defined as a triple, $O = (C,P,\textit{Axioms})$ where $C$ and $P$ correspond to the signature of $O$ and represent the classes and properties of $O$. The set \textit{Axioms} denotes a collection of axioms staying the main characteristics of the properties of $O$; these asserted statements implicitly comprise knowledge describing the modeled universe of discourse.
    \item [$\bullet$] The data sources of $DIS_G$ are represented by means of the set of signatures $S=\langle S_1^{A_1},\dots, S_n^{A_n} \rangle$ where each symbol $S_j$ stands for a data source, e.g., a file or relational table, and $A_j$ corresponds to the attributes of $S_j$: 
    \item [$\bullet$] The transformation of the data collected from the sources in $S$ into instances of the knowledge graph $G$ is expressed using the Global As View paradigm (GAV), i.e., the classes and properties in $O$ are described in terms of the sources $S$. The set $M$ comprises mapping rules $r_i$ where a class $c_j$ is described as a \emph{conjunctive query} on the sources and attributes in $S$. 
    \[r_i :  \underbrace{c_j(X,\overline{X})}_{\textit{Head of the Rule}} : - \underbrace{S_1(\overline{X_1}), S_2(\overline{X_2}),\dots, S_m(\overline{X_m})}_{\textit{Body of the Rule}} \]
    \begin{itemize}
    \item $c_j$ is a class in $C$, $X$ is a variable, and $\overline{X}$ is a set of pairs $(P_{i,j},X_{i,j})$ where $P_{i,j}$ is a property of $C$, i.e.,  $c_j$ is the domain of $P_{i,j}$, and $X_{i,j}$ is a variable. The variables $X_{i,j}$ and $X$ appear all in the body of the rule, i.e.,  $r_i$ is a \emph{safe conjunctive rule}. 
    \item The predicate $S_z(\overline{X_z})$ represents a source $S_z$ in $S$ and $\overline{X_z}$ is a set of pairs $(a_{i,z},X_{i,z})$ where $X_{i,z}$ is a variable and $att_{i,z}$ is an attribute of $S_z$, i.e., $S_z^{A_z}$ and $att_{i,z}$ belong to $S$ and $A_z$, respectively. 
    \end{itemize}
\end{itemize}

Given a data integration system $DIS_G = \langle O,S,M \rangle$, the evaluation of each of the rules $r_i$ in $M$ according to the data in the sources in $S$, generates the RDF knowledge graph $G$. The evaluation of $r_i$, $\textit{eval}(r_i,\mu)$, is defined over a map $\mu$ of the variables in $r_i$ to values in the sources in the body of $r_i$. A map $\mu$ corresponds to a function from variables $V$ in the rules in $M$ to the set $D$ which denotes the union of all the data items in the data sources in $S$, i.e., $\mu: V\rightarrow D$. 

Given a source predicate $S_z(\overline{X_z})$ in the body of a rule $r_i$, the evaluation of $S_z(\overline{X_z})$ on $\mu$, $\textit{eval}(S_z(\overline{X_z}),\mu)$, corresponds to a set $\mu_{S_z}$ of pairs, such that, for every pair $(att_{i,z}, X_{i,z})$ in $\overline{X_z}$, the following statements hold: 
\begin{itemize}
    \item  The pair $(X_{i,z}, \mu(X_{i,z}))$ belongs to $\mu_{S_z}$ and 
    \item If $\langle att_{1,z},\dots, att_{q,z} \rangle$ are the attributes of $S_z$ in $\overline{X_z}$, then the tuple 
    \[\langle (att_{1,z},\mu(X_{1,z})),\dots, (att_{q,z},\mu(X_{q,z}))\rangle \textrm{ belongs to the data extension of }S_z \]
\end{itemize}    

The evaluation of a rule $r_i$ on a map $\mu$, $\textit{eval}(r_i,\mu)$, 
corresponds to a set of RDF triples $t=(s \; p \; o)$ defined as follows: 
\begin{itemize}
    \item If the rule $r_i$ is $c_j(X,\overline{X}) : - S_1(\overline{X_1})$ and the pair $(X, \mu(X))$ belongs to $\mu_{S_1}$, then for each $(X_{i,1}, \mu(X_{i,1}))$ in $\mu_{S_1}$ and  $(P_{i,1},X_{i,1})$ in $\overline{X}$, the RDF triple
    $t=(\mu(X) \; P_{i,1}  \;\mu( X_{i,1}))$ belongs to  $\textit{eval}(r_i,\mu)$.
    \item Suppose the rule $r_i$ is $c_j(X,\overline{X}) : - \textit{Body}$, $\mu$ is defined over all the variables of the sources $S_Z$ in \textit{Body}, and
    the pair $(X, \mu(X))$ belongs to at least one $\mu_{S_z}$.  Then for each $(X_{i,z}, \mu(X_{i,z}))$ in $\mu_{S_z}$ and  $(P_{i,z},X_{i,z})$ in $\overline{X}$, the RDF triple
    $t=(\mu(X) \; P_{i,z}  \;\mu( X_{i,z}))$ belongs to  $\textit{eval}(r_i,\mu)$.
    
\end{itemize}    

Given a data integration system $DIS_G = \langle O,S,M \rangle$ the function \emph{RDFize(.)} maps $DIS_G$ with a knowledge graph $G$ resulting from the evaluation of all the rules in $M$ with the maps $\mu$ in the extensions of the sources in $S$. The result of the function \emph{RDFize(.)} only dependents on the mapping rules in $M$ and the extensions of the sources in $S$ over which these rules are evaluated. Nevertheless, in presence of data sources characterized with a large number of duplicates, the execution time of the function \emph{RDFize(.)} can be ngatively impacted. In this paper, we tackle the problem of rewriting a data integration system $DIS_G = \langle O,S,M \rangle$ into another data integration system   $DIS'_G = \langle O,S',M' \rangle$ whose evaluation produces the same results while the execution time is minimized.   

\begin{figure}[H]
\includegraphics[width=\textwidth]{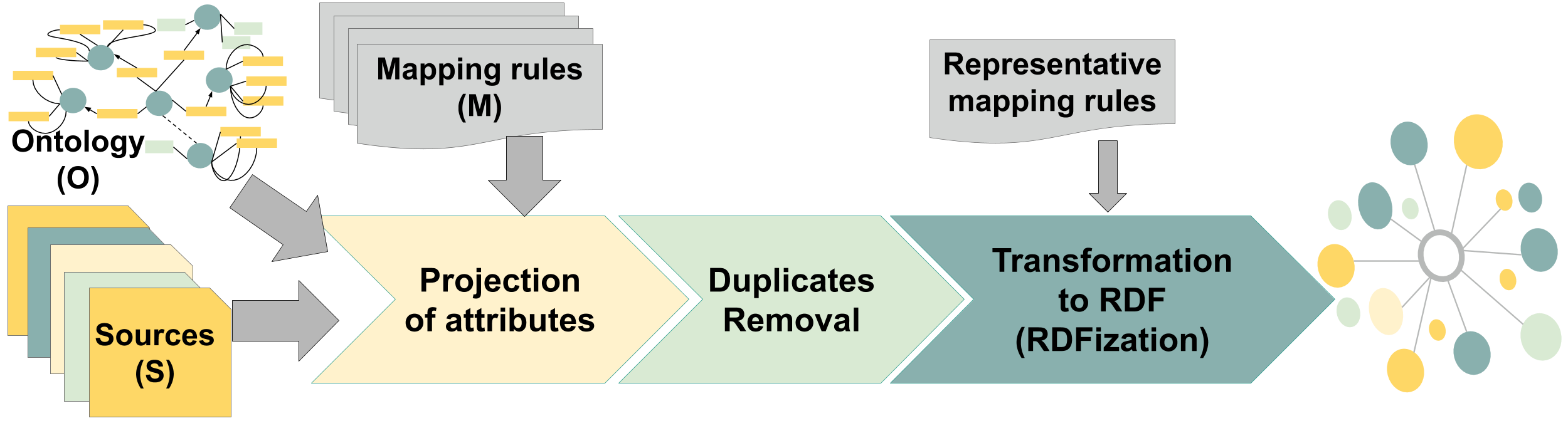}
\vspace{-0.5cm}
\caption{\textbf{The MapSDI framework.} MapSDI receives as input a data integration system and produces as output RDF triples to be included in a knowledge graph. MapSDI extracts from the mapping rules information related to the attributes that are used from each file. Then, different operations are executed to project out the required attributes; projection eliminates duplicates inside each dataset. Next, datasets comprising equivalent attributes are merged and duplicates are eliminated. The mapping rules are rewritten accordingly in order to access the transformed files, and finally, the mapping rules are executed.} \label{fig:MapSDI}
\end{figure}

\noindent \textbf{Problem Statement:}
Given a data integration system $DIS_G = \langle O,S,M \rangle$, the \emph{problem of knowledge graph creation} is defined as the problem 
of identifying a data integration system $DIS'_G = \langle O,S',M' \rangle$ such that:
\begin{itemize}
    \item The results of evaluating the two data integration systems is the same, i.e., $\emph{RDFize}(DIS_G = \langle O,S,M \rangle)$=$\emph{RDFize}(DIS'_G = \langle O,S',M' \rangle)$.
    \item The execution time of the evaluation of $\emph{RDFize}(DIS'_G = \langle O,S',M' \rangle)$ is  \emph{minimal}, i.e., there is no other $DIS''_G$ different from $DIS'_G$ that generates the same RDF knowledge graph $G$ but in a lower execution time. 
\end{itemize}    
\noindent \textbf{Proposed Solution:}
We propose MapSDI, an optimized alternative to traditional semantic data integration pipelines to create knowledge graphs. As it is shown in \autoref{fig:MapSDI}, MapSDI receives a data integration system $DIS_G = \langle O,S,M \rangle$ as input and generates an RDF knowledge graph that corresponds to the result of evaluating $\emph{RDFize}(DIS_G = \langle O,S,M \rangle)$. Without lost of generality, MapSDI assumes that the mapping rules in $M$ are represented in a mapping language, e.g., the RDF mapping language RML.

Before evaluating the function \emph{RDFize(.)}, MapSDI applies transformations to the sources in $S$ and the mapping rules in $M$ in order to generate a data integration system $DIS'_G = \langle O,S',M' \rangle$ 
that corresponds to a solution of the \emph{problem of knowledge graph creation}. MapSDI resorts to transformation rules applied to mapping rules and source depending on the attributes, variables, and sources that compose the mapping rules in $M$. That is, in a rule $r_i$, the attributes from the data sources in the \emph{Body} of $r_i$ are detected, and the corresponding sources in $S$ are transformed in order to have in $S'$ only data sources associated with the attributes utilized in the mapping rules. Accordingly, mapping rules are also rewritten with the aim of reusing the attributes of the sources in $S'$. By projecting out only the attributes required in the mapping rules, duplicates from the extensions of the sources are removed, avoiding thus, the generation of the same RDF triple multiple times during the evaluation of the function \emph{RDFize(.)}. Since only duplicates in the data sources are removed from the input, the resulting knowledge graph remains being the same, while the time of producing duplicated RDF triples is reduced. 
\subsection{Transformations Performed in the MapSDI framework}
\begin{figure}[t!]
 \centering
\includegraphics[width=0.8\columnwidth]{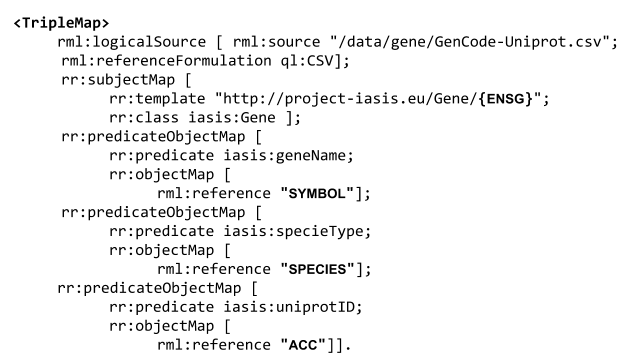}
    \caption{{\bf Example of Transformation Rule 1- Projection  of  Attributes}.  
    Only four attributes of a data source are utilized in the RML rule; processing the values of these four attributes conduce to many duplicated RDF triples. }
    \label{fig:tripleMap1}
\end{figure}

\begin{figure}[t!]
 \centering
 \subfloat[Portion of a Source File about Genes]{
      \includegraphics[width=0.9\columnwidth]{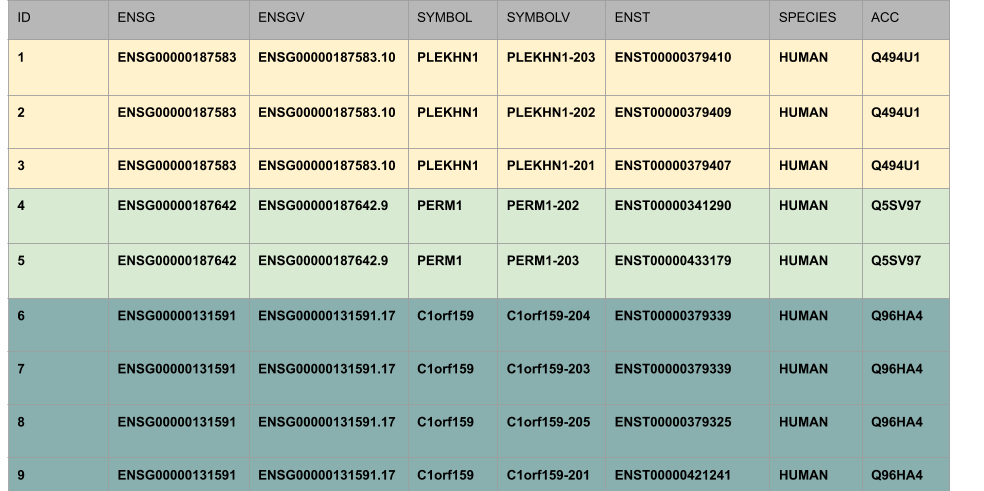}
    \label{fig:file1}}
    \\
  \subfloat[Source File After the Transformation Rule 1]{
\includegraphics[width=0.7\columnwidth]{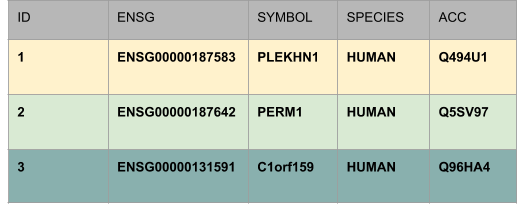}
    \label{fig:file1-2}}
    \caption{{\bf Example of Transformation Rule 1}. Projection of Attributes: (a) 
    RML Triple Map; only four attributes of the file are utilized in the rule; processing the values of these four attributes conduce to the generation of many duplicated RDF triples. (b)  A file with information about genes; several values are duplicates across the file.  (c) The file resulting of the projection of the attributes utilized in the triple map; the file does not have repeated  attributes and the execution of the triple map does not produce duplicated RDF triples.}
    \label{fig:rule1}
\end{figure}

We present the transformation rules applied by the MapSDI framework in order to reduce duplicated data and speed up the execution time of the evaluation of a data integration system. The transformation rules are based on the axioms of the relational algebra \cite{Ullman89} and in particular, the ones that stay when the project operator can be pushed down into the relations in a relational algebra expression. Furthermore, MapSDI extracts information from the mapping rules to decide when two or more datasets have equivalent attributes while represented with different attribute labels and must be merged into one file; and in case the merging is conducted, the corresponding rules are also merged. 
\\
\noindent \textbf{Transformation Rule 1: Projection of Attributes:}
A triple map may only use a subset of the attributes of a data source, generating thus high overhead whenever the number of attributes used in the triple map and the number of attributes in the data source differ considerably. To illustrate this situation consider the RML triple map in \autoref{fig:tripleMap1} whose evaluation produces many duplicates.
Additionally, the data source in Figure \autoref{fig:file1} comprises eight attributes but only four attributes are used in the rules. The values of the attributes \texttt{ENSG}, \texttt{SYMBOL}, \texttt{SPECIES}, and \texttt{ACC} are repeated, e.g., the rows 1,2, and 3 have the same values in these attributes, and similarly rows 4 and 5, and 6, 7, 8, and 9, respectively. Coincidentally, the evaluation of the triple map in \autoref{fig:tripleMap1} creates RDF triples from these four attributes and because during the execution of this triple map the data source is blindly traversed, several duplicated RDF triples are generated. \textit{Transformation rule 1} reduces the overhead caused whenever a triple map utilizes only a subset of the attributes of a data source; it pushes down the projection of the triple map object attributes before the triple map is executed. Thus, during the execution of the triple map only three rows are processed and no duplicated RDF triples are generated. In the case reported in \autoref{fig:rule1}, processing the original file in Figure \autoref{fig:file1} and the RML triple map (\autoref{fig:tripleMap1}) generate five duplicated RDF triples. Contrary, when file in Figure \autoref{fig:file1-2} is utilized, no duplicates are produced, thus the overhead during knowledge graph creation is considerably reduced. The time savings are reported in \autoref{sec:experiments}.

\begin{figure}[t!]
 \centering
 \includegraphics[width=1.0\columnwidth]{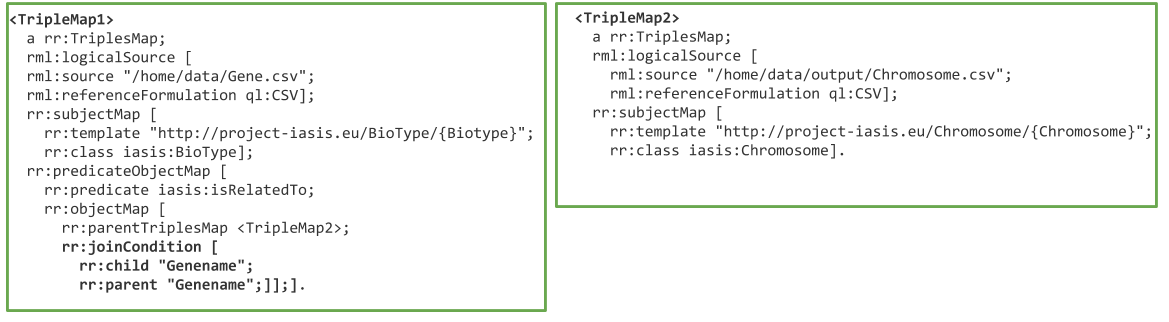}
 \vspace{-0.4cm}
 \caption{{\bf Example of RML Triple Maps}.Two RML Triple Maps connected by a join condition on the attribute \texttt{Genename}. Due to the number of duplicated values of the attribute \texttt{Genename}, the evaluation of the join condition generates a large number of duplicates. The projection of the relevant attributes reduce the number of duplicated values and RDF triples.}.
    \label{fig:tripleMap2}
    \vspace{-1.0cm}
\end{figure}
  
\begin{figure}[t!]
 \centering
 \subfloat[Portion of a Source File about Genes (Outer Source File)]{
      \includegraphics[width=0.9\columnwidth]{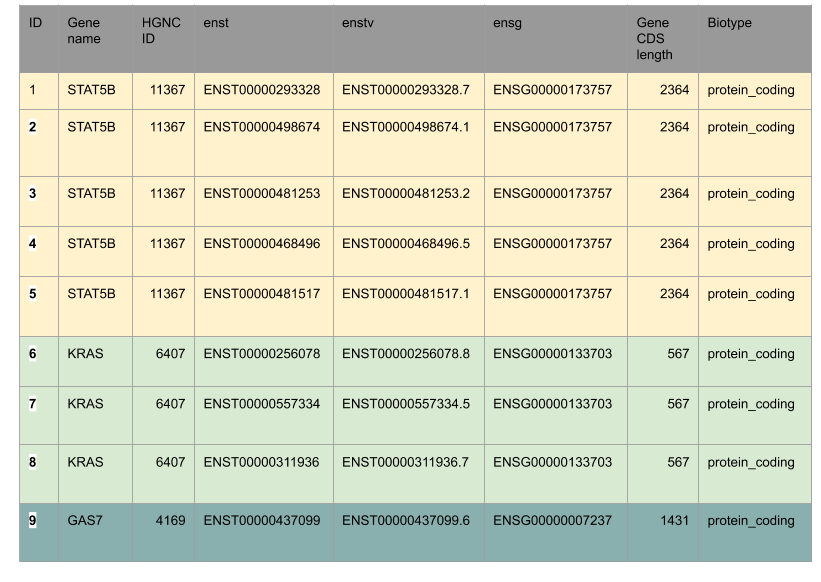}
    \label{fig:file2}}
    \\
   \vspace{-0.4cm}
  \subfloat[Portion of the Source File about Chromosomes (Inner Source File)]{
\includegraphics[width=0.9\columnwidth]{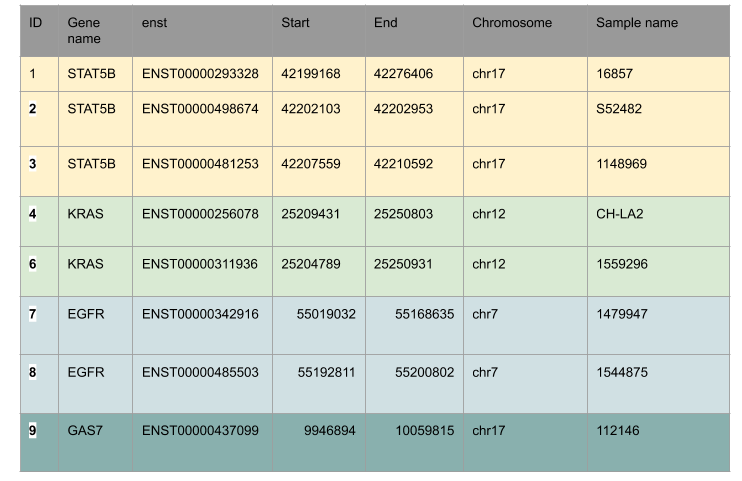}
    \label{fig:file2-2}}
    \caption{{\bf Example of Transformation Rule 2}. Pushing down Projections into a Join: (a) and (b)  Files containing data to be considered as the outer and inner data sources of \texttt{TripleMap1} (\autoref{fig:tripleMap2}), respectively. Duplicates in the join attribute  conduce the generation of 22 duplicated RDF triples.}
    \label{fig:rule2}
\end{figure}

\begin{figure}[t!]
 \centering
 \subfloat[Projection on Genes]{
      \includegraphics[width=0.45\columnwidth]{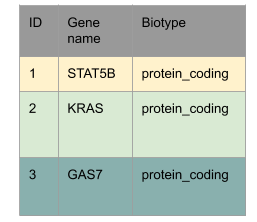}
    \label{fig:file3}}
  \subfloat[Projection on Chromosomes]{
\includegraphics[width=0.45\columnwidth]{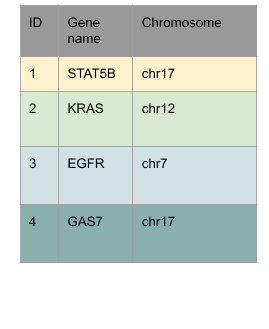}
    \label{fig:file4}}
    \\
   \vspace{-0.4cm}
  \subfloat[RDF triples with reduced duplicates]{
\includegraphics[width=0.5\columnwidth]{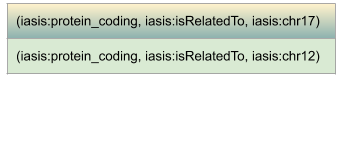}
    \label{fig:duplicates}}
    \caption{{\bf Example of Transformation Rule 2}. Pushing down Projections into a Join: (a) and (b) Projecting out from files in Figures \autoref{fig:file2} and  \autoref{fig:file2-2} the attributes mentioned in triple maps in  Figure \autoref{fig:tripleMap2}. (c) RDF triples produced by the triple maps over the projected attributes; duplicates are reduced from 22 to four.}
    \label{fig:rule2b}
\end{figure}

\noindent \textbf{Transformation Rule 2: Pushing Down Projection into Joins:}
This rule is applied whenever a join exists between two triple maps $r_1$ and $r_2$ defined over data sources with a large number of attributes that are not utilized in $r_1$ and $r_2$. To illustrate this case, consider \autoref{fig:tripleMap2}; the triple maps \texttt{TripleMap1} and \texttt{TripleMap2} are joined by the join condition highlighted in bold in \texttt{TripleMap1}. When this join is executed on datasets in Figures \autoref{fig:file2} and \autoref{fig:file2-2}, 22 duplicated RDF triples are generated. Duplicate generation considerably impacts on the performance of a knowledge graph creation, particularly, whenever duplicates are blindly generated and then, eliminated. To reduce the effect of duplicates during the evaluation of join conditions between two triple maps, MapSDI pushes the projections of the relevant attributes down before the triple maps are executed. As observed in \autoref{fig:rule2b}, this transformation considerably reduces the number of matches of the join condition and the resulting RDF triples. 

Once the attributes mentioned in the triple maps in \autoref{fig:tripleMap2} are projected out (files in Figures \autoref{fig:file3} and \autoref{fig:file4}), the execution of these triples maps still produces RDF triples that are duplicated. However, the number of duplicates is reduced from 22 to four. Considerably reducing thus, the workload required to generate, check, and eliminate duplicated RDF triples. Results of the experimental study will show the improvements of the MapSDI framework.  

\noindent \textbf{Transformation Rule 3: Merging data sources with equivalent attributes:} This rule is applied whenever there exist two or more triple mapping rules that generate the same type of subjects associated with the same predicates, but the data is collected from different data sources with attributes that may have different names. This rule allows the MapSDI framework to first, project the relevant attributes, and then merge the data sources; duplicates are eliminated from the merged data source. Additionally, the triple maps are merged in one triple map that will access the merged data source and duplicated RDF triples are not generated (See \autoref{fig1}).

MapSDI applies the transformation rules 1-3 over the input data integration system $DIS_G = \langle O,S,M \rangle$ in order to generate $DIS'_G = \langle O,S',M' \rangle$; these rules are applied until a fixed point over $S'$ and $M'$ is reached. 

\subsection{Correctness of the Transformation Rules}
\label{sec:formalAnalysis}
We demonstrate the correctness of the transformation rules 1-3 by proving that the application of each of these rules preserves the set of RDF triples produced during the evaluation of the original data integration system; these proofs are grounded on the axiomatic system of the Relational Algebra \cite{silberschatz1997database}.
\\ \\
\noindent \textbf{Transformation Rule 1: Projection of Attributes}.
\label{sec:transRule1}
For each mapping rule $r_i$ in $M$ with sources $S_z(\overline{X_z})$ in the body of $r_i$, the transformation rule 1, adds new sources  $S'_z$ to 
$S'$, in the way, that $S'_z$ is equal to $\prod_{Att}S_z$ and $Att$ is the set of attributes utilized in $\overline{X_z}$. The rule $r_i$ is removed from $M'$ and a new mapping rule  $r'_i$ where all the sources $S_z(\overline{X_z})$ are replaced by $S'_z(\overline{X_z})$. Since the attributes from the sources $S_z$ used in $\overline{X_z}$ are maintained in the new data sources $S'_z$ and in the rule $r'_i$, the results of $\emph{RDFize}(DIS_G = \langle O,S,M \rangle)$ and $\emph{RDFize}(DIS'_G = \langle O,S',M' \rangle)$ are the same. 
\\ \\
\noindent \textbf{Transformation Rule 2: Pushing  Down  Projection  into  Joins}.
Transformation rule 2 is applied over a mapping rule $r_i$ whenever there exist attributes and variables in the sources of the body of $r_i$ that are not required to evaluate $r_i$, i.e., they are neither used to instantiate the head of $r_i$ nor to join two or more data sources in the body of $r_i$. If so, transformation rule 2 projects out from the sources $S_z(\overline{X_z})$ in the body of $r_i$ the attributes and variables that are required. 
Formally, the rewriting of $r_i$ is defined as follows:
Let $\overline{Z}$ be the set of variables in the head of $r_i$ or in the join of at least two sources in the body. That is,
 $\overline{Z}$ is the union of variables in $\overline{X}$, $X$, and the variables that appear in more than one  $S_p(\overline{X_p})$ and $S_q(\overline{X_q})$ in the body of $r_i$.

\[r_i: c_j(X,\overline{X}):- S_1(\overline{X_1}),S_2(\overline{X_2}) \dots S_m(\overline{X_m})\]

The application of the transformation rule 2, replaces $r_i$ by the rule $r'_i$:
\[r'_i: c_j(X,\overline{X}):- S_1(\overline{X'_1}),S_2(\overline{X'_2}) \dots S_m(\overline{X'_m})\]
 where each $X'_j$, $1 \leq j \leq m$, is defined as follows:
 \[X'_j = X_j - \{(att_{i,j},X_{i,j}) \mid  (att_{i,j},X_{i,j}) \in X_j \; and \; X_{i,j} \notin \overline{Z}\}\]

The transformation 2 is grounded on the axiomatic system of the Relational Algebra, specifically,  on the rule axiom that states the properties of distributing the Project operator over a Join (rule number 8 in \cite{silberschatz1997database}). Thus, after applying this transformation rule and replacing $r_i$ by $r'_i$ in $M'$, the results of $\emph{RDFize}(DIS_G = \langle O,S,M \rangle)$ and $\emph{RDFize}(DIS'_G = \langle O,S',M' \rangle)$ are the same. 
\\ \\
\noindent \textbf{Transformation Rule 3: Merging data sources with equivalent attributes}.
This rule is applied over two mapping rules, $r_i$ and $r_j$, whenever both rules share the same head but the bodies are composed of different data sources, i.e., $r_i: c_q(X,\overline{X}):- S_i(\overline{X_i})$ and $r_j: c_q(X,\overline{X}):- S_j(\overline{X_j})$.
The result of applying the transformation rule 3 is a new data source $S_{i,j}$ that is populated with values of the attributes from  $S_{i}$ and
 $S_{j}$ that are required for instantiating $c_q(X,\overline{X})$.  Further, $r_i$ and $r_j$ are replaced by the rule $r_{i,j}$ in $M'$, $r_{i,j}: c_q(X,\overline{X}):- S_{i,j}(\overline{X_{i,j})}$
\begin{itemize} 
\item $S_{i,j}$ is the union of $\prod_{Att_i}S_i$ and $\prod_{Att_j}S_j$ such that $Att_i$ and $Att_j$, respectively,  are the attributes in $\overline{X_i}$ and $\overline{X_j}$ related with variables in $c_q(X,\overline{X})$. 
\item The projected attributes in $S_{i,j}$ are renamed and these new attributes are used in $\overline{X_{i,j}}$ associated with the corresponding variables in $c_q(X,\overline{X})$.
\end{itemize}

 The transformation 3 is also supported on the axiomatic system of the Relational Algebra, specifically,  on the rule axiom that states the properties of distributing the Project operator over a Union (rule number 12 in \cite{silberschatz1997database}). Thus, after applying this transformation rule and replacing $r_i$ and  $r_j$ by $r_{i,j}$ in $M'$, and adding the data source $S_{i,j}$ to $S'$, the results of $\emph{RDFize}(DIS_G = \langle O,S,M \rangle)$ and $\emph{RDFize}(DIS'_G = \langle O,S',M' \rangle)$ are the same.

\section{Experimental Study}
\label{sec:experiments}
We compare the performance of MapSDI to the traditional framework for knowledge graph creation which we refer to as "T-framework" from now on in this paper. We aim to answer the following questions: 
\begin{inparaenum}[\bf {\bf Q}1\upshape)]
\item Does applying MapSDI lead to creation of the same knowledge graph?  
\item Does MapSDI reduce the required time for knowledge graph creation compared to T-framework?
\item How influential is the performance of MapSDI framework, when data volume increases or data quality decreases?.
\item Does MapSDI perform efficiently in case of having more complication in mapping rules e.g., join condition? 
\end{inparaenum}
We set up the following testbeds:
\noindent \textbf{Datasets}
To prevent any bias that may arise due to using a specific database or data generated by a particular lab, several datasets have been combined. For the first experimental scenario, a dataset with an overall size of 312,1MB and 19,503,200 records is created from the combination of three different datasets including mutations, drug-resistant mutations, and protein-RNA interaction predictions; they are collected from different data providers: 
\begin{inparaenum}[\bf(i\upshape)]
\item The datasets related to mutations and drug-resistant mutations are collected from COSMIC\footnote{\url{https://cancer.sanger.ac.uk/cosmic}}, an open source database of somatic mutations in human cancer diseases. 
\item A dataset defined by Lang et al. \cite{lang2018rnact} at CRG\footnote{\url{https://www.crg.eu/}}, this dataset includes protein-RNA interaction predictions.
\end{inparaenum}
\noindent
The second studied dataset is generated by collecting different attributes from various publicly available datasets including the GENCODE reference annotation for the human and mouse genomes \cite{frankish2018gencode}. In this dataset, a large amount of selected data relates to exon, the sequence represented in the mature RNA whose mutations can directly affect the sequence of a protein \cite{principles}. Since there are overlaps between the data in these datasets, as we will explain later, there exist a large number of duplicates.

\begin{figure}[t!]
 \centering
    \subfloat[rmlmapper - 75\% veracity]{
        \includegraphics[width=0.45\columnwidth]{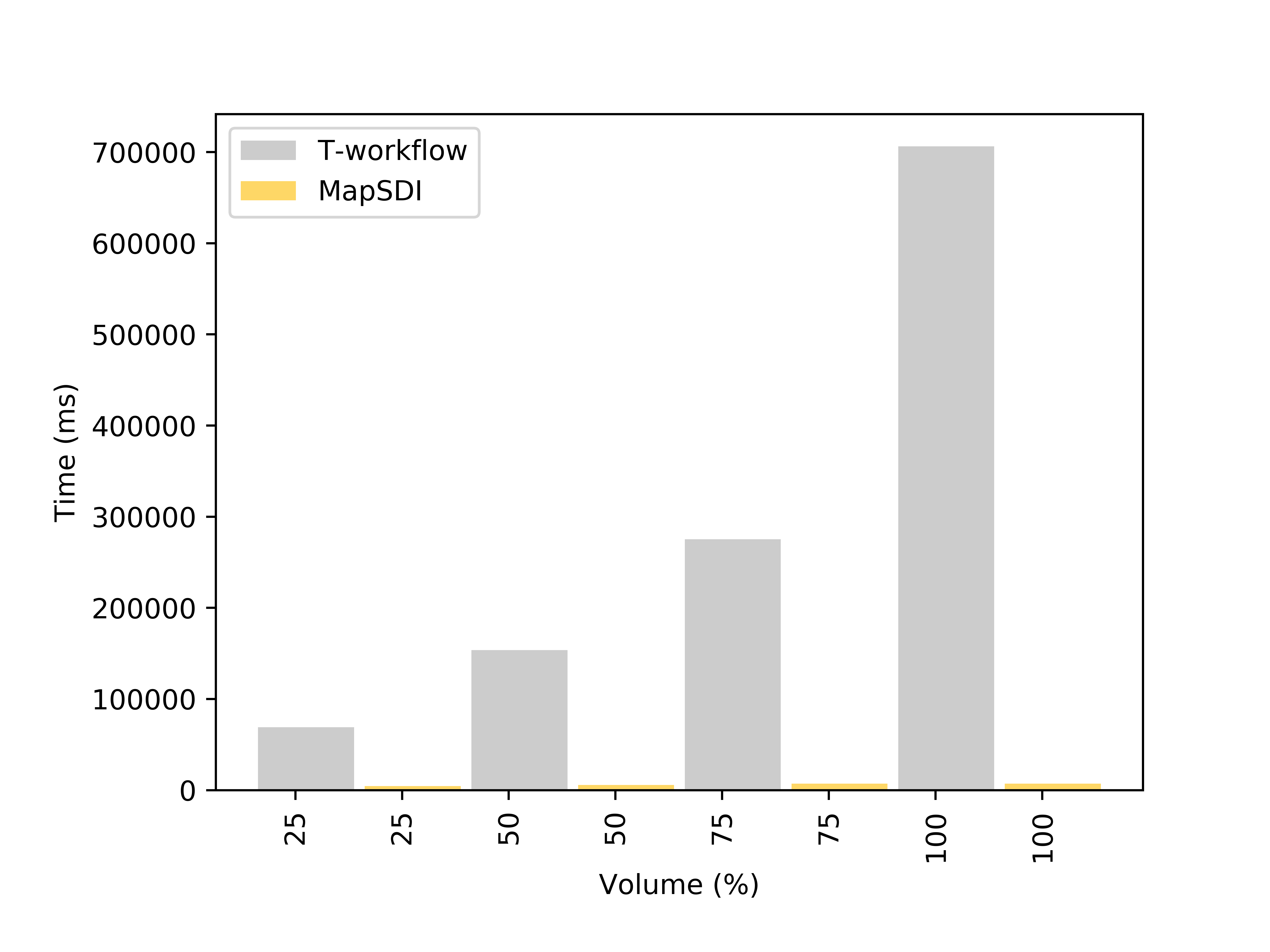}
            \label{fig:a_vera75_rmlMapper}}
    \subfloat[SDM-RDFizer - 75\% veracity]{
        \includegraphics[width=0.45\columnwidth]{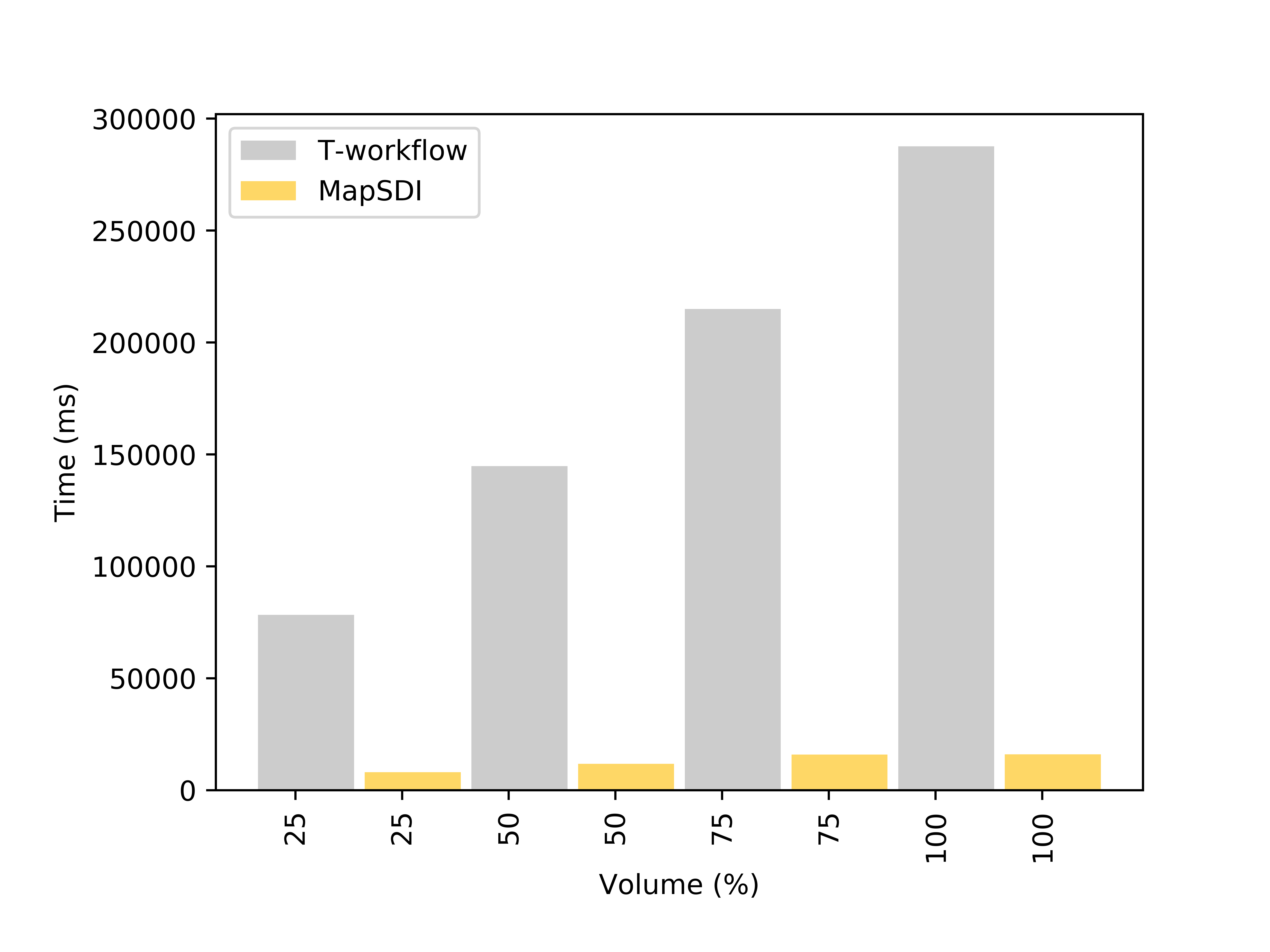}
            \label{fig:b_vera75_SDMrdfizer.png}}
\\            
    \subfloat[rmlmapper - 50\% veracity]{
        \includegraphics[width=0.45\columnwidth]{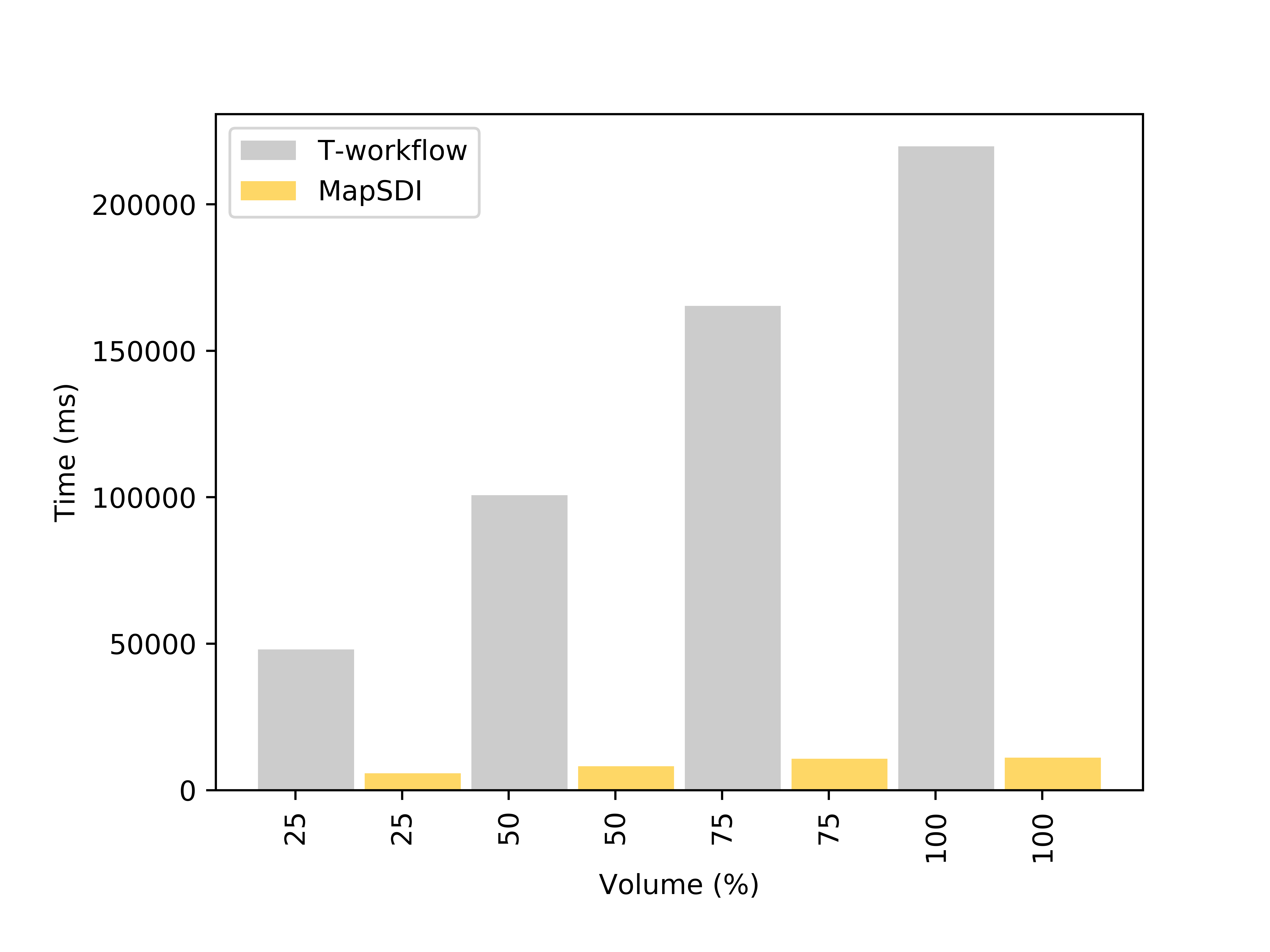}
            \label{fig:c_vera50_rmlMapper}}  
    \subfloat[SDM-RDFizer - 50\% veracity]{
        \includegraphics[width=0.45\columnwidth]{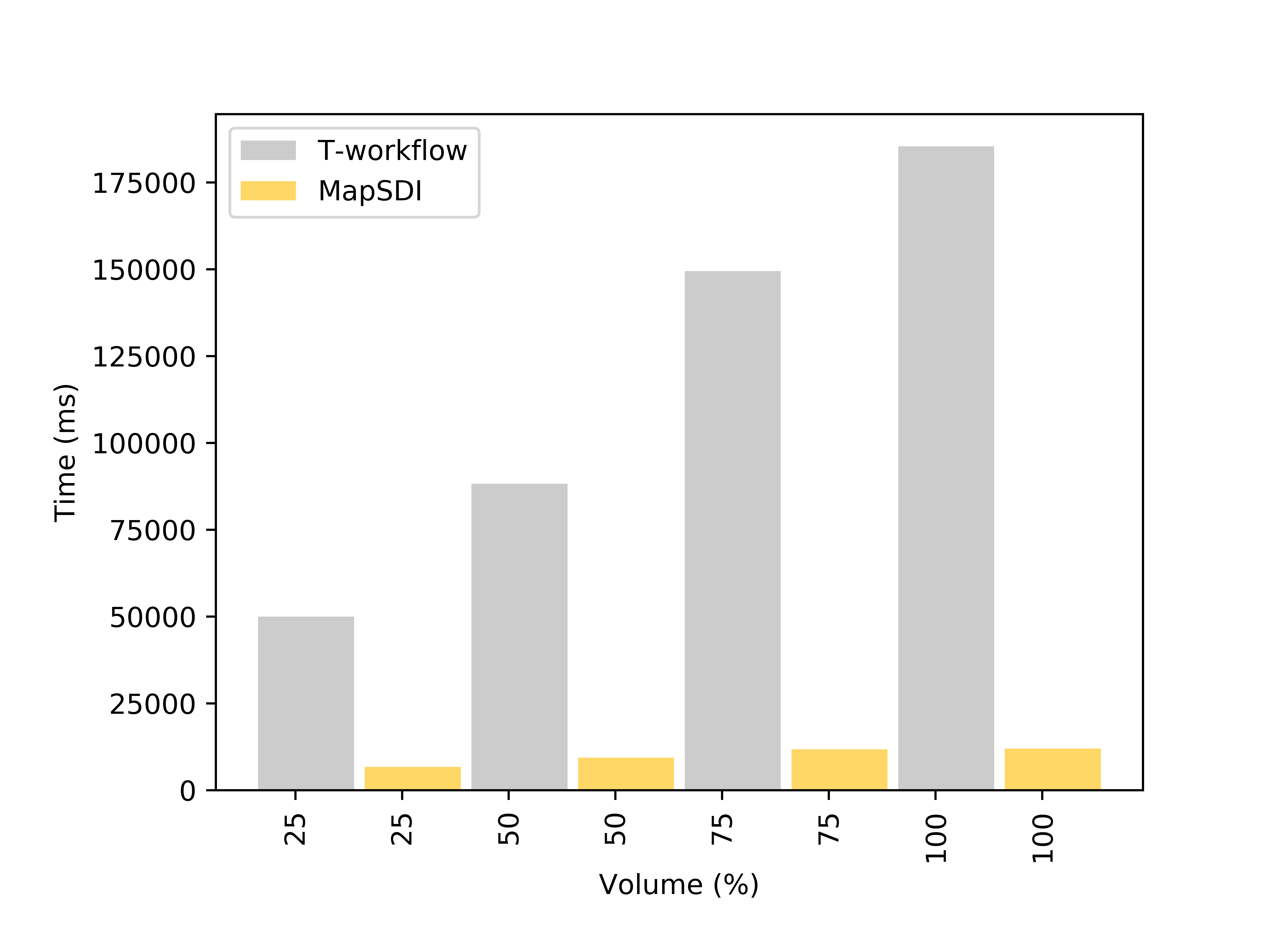}
            \label{fig:d_vera50_SDMrdfizer}} 
\\            
    \subfloat[rmlmapper - 25\% veracity]{
        \includegraphics[width=0.45\columnwidth]{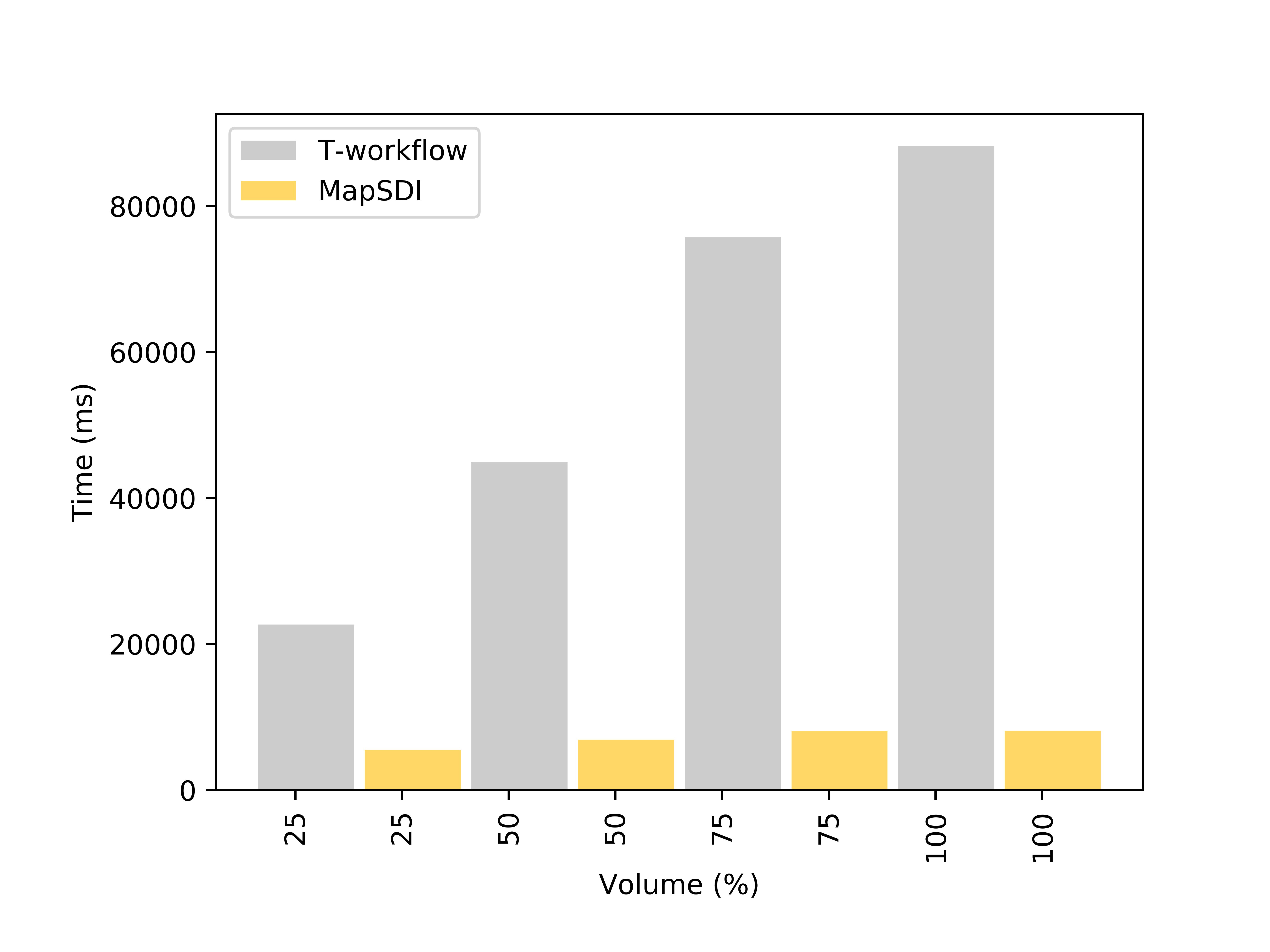}
            \label{fig:f_vera25_SDMrdfizer}}
    \subfloat[SDM-RDFizer - 25\% veracity]{
        \includegraphics[width=0.45\columnwidth]{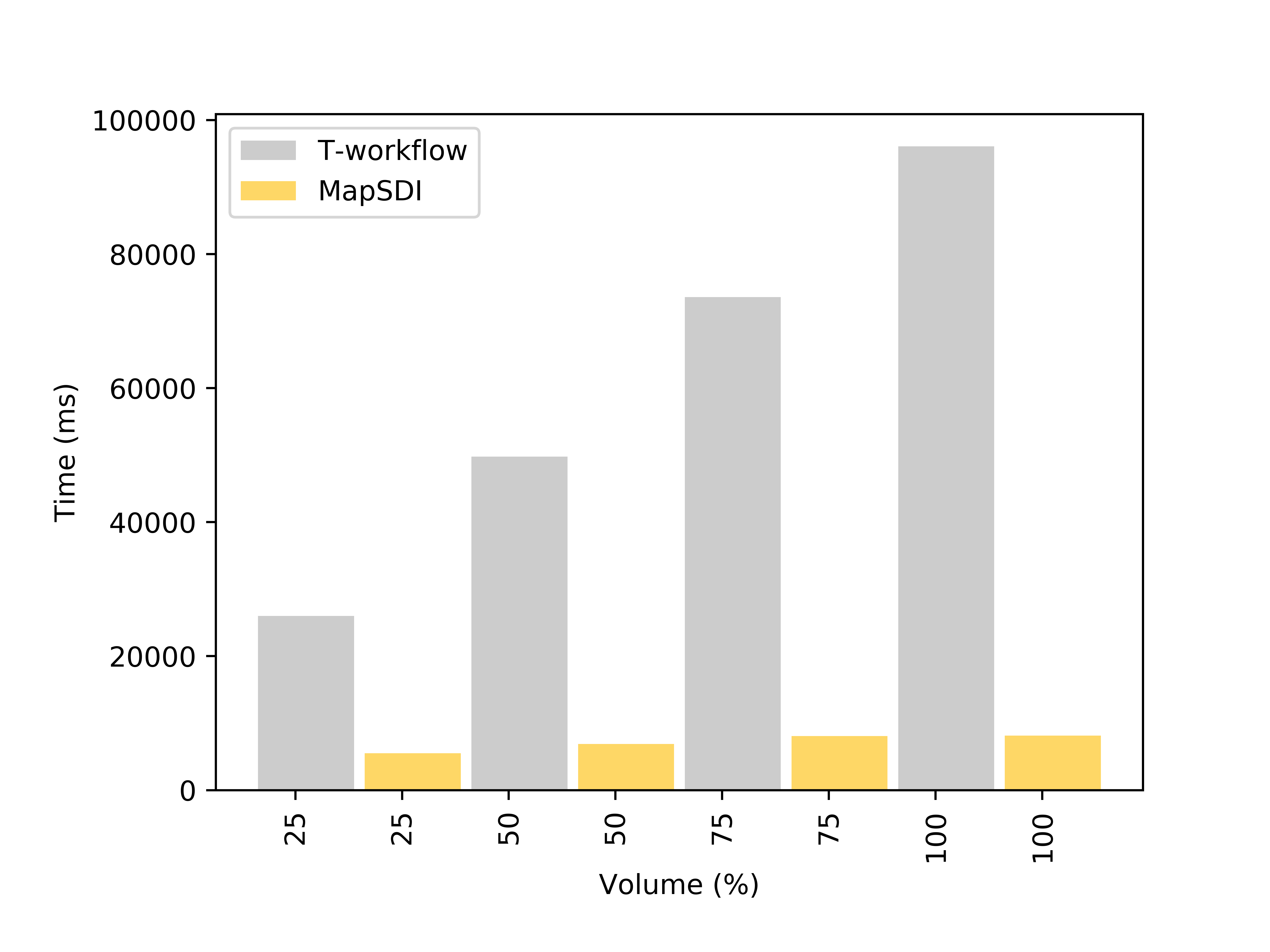}
            \label{fig:f_vera25_SDMrdfizerr}}              
\caption{{\bf Results of experiment group A with different percentage of veracity}. The performance of MapSDI and T-framework on four different sized datasets with 75\% redundancy: (a) applying rmlmapper (b) using SDM-RDFizer. MapSDI is able to reduce duplicated and exhibits better performance independently of the data volume and RDFizer. But, the difference between the execution time of two frameworks is much higher when rmlmapper is evaluated.}
    \label{fig:expA}
\end{figure}

\noindent \textbf{Metrics}
Performance is measured in terms of execution time; it is computed as the elapsed time in seconds between the submission of an execution of the framework and the generation of all the RDF triples. The \texttt{time} command of the Linux operating system is utilized to measure time. The timeout is set to 500 seconds; the results are visualized based on milliseconds.  

\noindent \textbf{Implementations}
MapSDI and T-framework are compared on 
SDM-RDFizer\footnote{\url{https://github.com/SDM-TIB/TIB-RDFizer}}\linebreak and the rmlmapper-java\footnote{\url{https://github.com/RMLio/rmlmapper-java}}. The MapSDI framework is implemented in Python 3.6.3 and GNU bash 4.4.12(1) jointly. The experiments are executed on an Ubuntu 17.10 (64 bits) machine with Intel Xeon W-2133, CPU 3.6GHz, 1 physical processor; 6 cores, 12 threads and 64 GB RAM. 

\noindent \textbf{Experimental Scenarios}
We perform in overall 51 experiments; divided into two groups of studies. 
\begin{inparaenum}[{\bf Group} A \upshape)]
\item
The first group of experiments are designed to study the impact of the size of input datasets and their quality in terms of redundancy, on required time for semantic enrichment and integration. In order to avoid the experiments being influenced by other variables such as the number of included attributes and mapping rules, in all experiments of this group, the same one concept is utilized; this concept is represented as a different attribute in each dataset. Additionally, to highlight the difference between the performance of two frameworks, a minimal setup consisting of one attribute in each dataset and consequently one RML triple map, are evaluated.
Each 12 experiments that are performed based on a separated framework using a different RDFizer, can be divided into four categories based on the data volume: 
the \textbf{25\%}, \textbf{50\%}, \textbf{75\%}, and \textbf{100\% volume}; they are produced by randomly selecting 25\%, 50\%, 75\% and 100\% of the records in created dataset, respectively. Subsequently, each mentioned category is divided into three subcategories based on data redundancy; from each generated dataset in the volume category, three datasets are produced by cleaning 25\%, 50\% and 75\% of the data from duplicates. It should be noted that all selections of data have been performed randomly to avoid any sampling bias.   
\item
The second experiment setup is conducted to study the impact of data redundancy on performance of each framework in case of join condition rules inclusion. Following the same objective, the minimum amount of required attributes are considered. Accordingly, three experiments are performed on joining two datasets: 
\begin{inparaenum}[a\upshape)]
\item No dataset with duplicates removal;
\item One dataset being duplicates-free; and
\item Both datasets being duplicates-free.
\end{inparaenum}
\end{inparaenum}

\begin{table}[t!] 
\caption{\textbf{Four instance datasets size}. The size of four datasets applied in experiments group A with the results being shown in \autoref{fig:expA}. The values show how the size of datasets are reduced after the two steps of attribute projection and duplicate removal have been applied on, as part of MapSDI framework.}
\label{tab:1}
\centering
\begin{tabular}{|c|c|c|}
\hline
\textbf{Data Volume} & \textbf{Original Size (KB)} & \textbf{Pre-processed Size (KB)} \\
\hline
25\% & 59,200 & 895\\
\hline
50\% & 117,900 & 955\\
\hline
75\% & 176,400 & 982 \\
\hline
100\% & 235,000 & 997 \\
\hline
\end{tabular}
\end{table}

\begin{figure}[t!]
 \centering
\includegraphics[width=0.7\columnwidth]{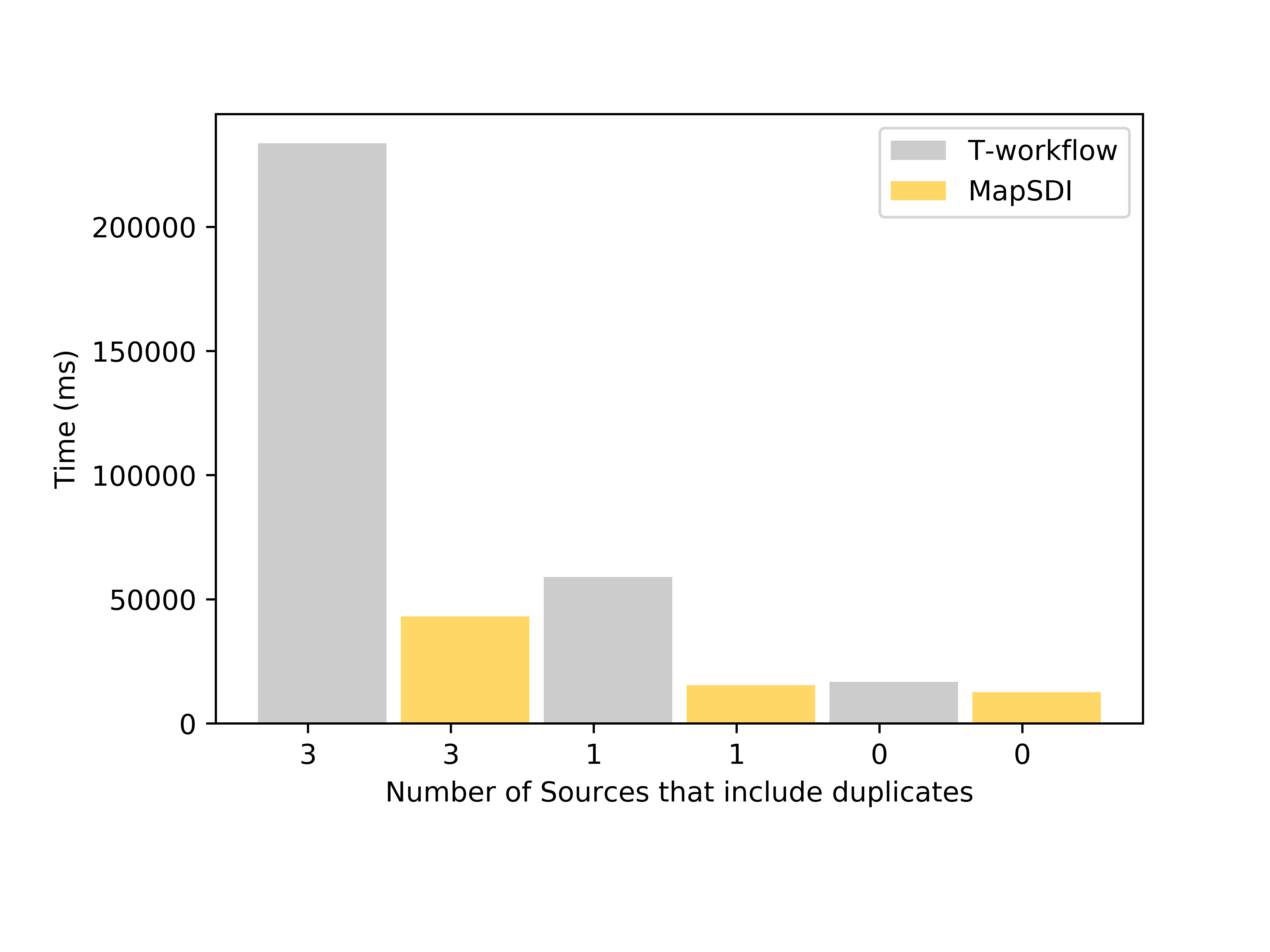}
\vspace{-0.1cm}
\caption{{\bf Results of Experiment Group B}. MapSDI and T-framework on two datasets joined by two triple maps. MapSDI performs \textit{Transformation Rule 2 and Rule 3} and it is able to push down projection into the join. With the transformations conducted by MapSDI, the rmlmapper timed out at 500 seconds.}
 \label{fig:expB}
\end{figure}
\subsection{Experimental Results}
\noindent \textbf{Experimental results group A:}
The results of the experiment group A are shown in Figures \ref{fig:expA}. As it can be observed, MapSDI outperforms T-framework in terms of execution time in all the experiments independently of the RDFizers and percentage of duplicates. This instance of the MapSDI framework performs the \textit{Transformation Rule 3}, i.e., the datasets are merged; while the \textit{Transformation Rule 1} is performed in the two frameworks during the creation of the datasets. According to the results depicted in Figures \ref{fig:expA}, regardless of the RDFizer, the more duplicated data in the datasets, the higher the execution time of the T-framework. It is also important to highlight, the diverse performance ratios of MapSDI and T-framework in terms of the growth of dataset size and data duplicates. MapSDI performs more stable than T-framework. These observations can be explained according to the two steps of pre-processing including attributes projection and duplicates removal that are executed former to the transformation step in the MapSDI framework. The mentioned steps decrease the size of the original datasets considerably. \autoref{tab:1} reports on the reduced size of the input datasets after the pre-processing steps in the experiments conducted over the dataset with 25\% data duplicates (\autoref{fig:expA}).  
\\ \\
\noindent \textbf{Experimental results group B:}
\autoref{fig:expB} illustrates the results of experiments in group B. The rmlmapper timed out in all experiments of group B, the results only refer to the performance of MapSDI and T-framework applying SDB-RDFizer. As it can be observed, the execution time of MapSDI is considerably lower than T-framework in case of having join condition in mapping rules independent of having data duplicates. This instance of MapSDI framework performs the \textit{Transformation Rule 3} as well as \textit{Transformation Rule 2}. The application of these two transformations considerably reduces the number of duplicates and enhances the performance of the SDM-RDFizer during the execution of the join condition between two triple maps. 
\section{Related Work}
\label{sec:relatedWork} 
The problem of knowledge graph creation is one of the trending topics which also involves different problems such as data integration. Lenzerini et al \cite{lenzerini2002data} provides an overview on the components required to define a data integration system. Gawriljuk, et al. \cite{gawriljuk2016scalable} suggest a scalable framework for building knowledge graphs. Szekely, et al. \cite{szekely2015building} propose an approach for building knowledge graphs and devise the DIG system which resorts to KARMA \cite{knoblock2015exploiting}, a semantic data integration system proposed by Knoblock et al., for integration at the level of schema. Collarana et al. introduce MINTE \cite{collarana2017minte}, a semantic integration technique for RDF graphs. Although the mentioned approaches are effective, they either differentiate between the integration at the level of schema and the data-level integration or only focus on one of the two tasks. This distinction leads to a dramatic increase in the cost of semantic data integration in case of consuming big data. In contrast, in MapSDI both integration tasks are conducted  simultaneously. Moreover, the semantics encoded in the schema and mapping rules is utilized in order to first, remove the data redundancy and then, transform the input data into RDF triples. 
Diverse mapping languages for transforming relational data into RDF have been introduced, reported in 2009 for the first time as a survey by W3C incubator group. Sequeda et al. explain the limitations of semantic technologies in relational databases integration in \cite{sequeda2017integrating}. During the recent years several extension to R2RML have been proposed in order to represent mapping rules such as RML \cite{dimou2013extending} by Dimou et al. or D2RML \cite{chortaras2018d2rml} by Chortaras et al. The same applies for the implementation of tools to execute mapping rules in different languages. In this work, we present MapSDI, a framework that is able to speed up the execution time of the task of knowledge graph creation independently of the mapping language or tools for knowledge graph creation. Experimentally, we have observed that MapSDI empowers the performance of the RDFizers regardless of the number of duplicates and size of the input data.  
\section{Conclusions and Future Work}
\label{sec:conclusions}
We tackled the problem of optimizing semantically integrating data into a knowledge graph and presented MapSDI; it is devised for enabling the semantic enrichment of data characterized by the dominant dimensions of big data, i.e., volume, variety, and veracity. MapSDI resorts to the properties of the relational algebra operators and to the knowledge encoded in the mapping rules to identify the transformations that need to be performed to the input data to empower the performance of existing knowledge graph creation tools. Thus, our resource broadens the repertoire of techniques available to integrate heterogeneous datasets into a knowledge graph, and we hope that these techniques will help the community in the development of more scalable knowledge graph based applications. In the future, we will extend the MapSDI framework to include other transformations and mapping languages. Furthermore, the development of applications on top of the MapSDI framework is part of our future plans. 
\paragraph{\bf Acknowledgements} This work has been partially funded by the EU H2020 Program for the Project No. 727658 (IASIS).

\bibliographystyle{abbrv}
\bibliography{bibliography}

\begin{thebibliography}{10}

\bibitem{antoniou2004semantic}
G.~Antoniou and F.~Van~Harmelen.
\newblock {\em A semantic web primer}.
\newblock MIT press, 2004.

\bibitem{chortaras2018d2rml}
A.~Chortaras and G.~Stamou.
\newblock D2rml: Integrating heterogeneous data and web services into custom
  rdf graphs.
\newblock In {\em LDOW@ WWW}, 2018.

\bibitem{collarana2017minte}
D.~Collarana, M.~Galkin, I.~Traverso-Rib{\'o}n, M.-E. Vidal, C.~Lange, and
  S.~Auer.
\newblock Minte: semantically integrating rdf graphs.
\newblock In {\em Proceedings of the 7th International Conference on Web
  Intelligence, Mining and Semantics}, 2017.

\bibitem{dimou2013extending}
A.~Dimou, M.~Vander~Sande, P.~Colpaert, E.~Mannens, and R.~Van~de Walle.
\newblock Extending r2rml to a source-independent mapping language for rdf.
\newblock In {\em International Semantic Web Conference (Posters \& Demos)},
  volume 1035, 2013.

\bibitem{frankish2018gencode}
A.~Frankish, M.~Diekhans, A.-M. Ferreira, R.~Johnson, I.~Jungreis, J.~Loveland,
  J.~M. Mudge, C.~Sisu, J.~Wright, J.~Armstrong, et~al.
\newblock Gencode reference annotation for the human and mouse genomes.
\newblock {\em Nucleic acids research}, 47(D1), 2018.

\bibitem{gawriljuk2016scalable}
G.~Gawriljuk, A.~Harth, C.~A. Knoblock, and P.~Szekely.
\newblock A scalable approach to incrementally building knowledge graphs.
\newblock In {\em International Conference on Theory and Practice of Digital
  Libraries}, 2016.

\bibitem{knoblock2015exploiting}
C.~A. Knoblock and P.~Szekely.
\newblock Exploiting semantics for big data integration.
\newblock {\em AI Magazine}, 36(1), 2015.

\bibitem{lang2018rnact}
B.~Lang, A.~Armaos, and G.~G. Tartaglia.
\newblock Rnact: Protein--rna interaction predictions for model organisms with
  supporting experimental data.
\newblock {\em Nucleic acids research}, 47(D1), 2018.

\bibitem{lenzerini2002data}
M.~Lenzerini.
\newblock Data integration: A theoretical perspective.
\newblock In {\em Proceedings of the twenty-first ACM SIGMOD-SIGACT-SIGART
  symposium on Principles of database systems}. ACM, 2002.

\bibitem{principles}
B.~Lewin, J.~Krebs, S.~T. Kilpatrick, and E.~S. Goldstein.
\newblock {\em Lewin's GENES X}.
\newblock 2011.

\bibitem{sequeda2017integrating}
J.~F. Sequeda.
\newblock Integrating relational databases with the semantic web: A reflection.
\newblock In {\em Reasoning Web International Summer School}, 2017.

\bibitem{silberschatz1997database}
A.~Silberschatz, H.~F. Korth, S.~Sudarshan, et~al.
\newblock {\em Database system concepts}, volume~4.
\newblock McGraw-Hill New York, 1997.

\bibitem{szekely2015building}
P.~Szekely, C.~A. Knoblock, J.~Slepicka, A.~Philpot, A.~Singh, C.~Yin,
  D.~Kapoor, P.~Natarajan, D.~Marcu, K.~Knight, et~al.
\newblock Building and using a knowledge graph to combat human trafficking.
\newblock In {\em International Semantic Web Conference}, 2015.

\bibitem{Ullman89}
J.~D. Ullman.
\newblock {\em Principles of Database and Knowledge-Base Systems, Volume {II}}.
\newblock Computer Science Press, 1989.

\end{thebibliography}
\end{document}